\documentclass[aps,prl,final,twocolumn,letterpaper]{revtex4}

\usepackage{graphicx}   
\usepackage{import}                         % Graphics package
\usepackage{yhmath}
\usepackage{epstopdf}
\usepackage{amsmath} 
\usepackage{float} 
\usepackage{bm}
\usepackage{amssymb}
\usepackage{quotes}
\usepackage{indentfirst}
\usepackage{color}
\usepackage{transparent}
\usepackage{dcolumn}% Align table columns on decimal point
\usepackage{braket}
\usepackage{multirow}
\usepackage{cancel} % for the command \cancel and \cancelto
\usepackage{mdframed}
\usepackage{color}
\usepackage{bm}
\usepackage{dsfont}
\usepackage{slashed}
\usepackage{soul, color} % for the \hl and \st commands
\soulregister\ref{7}  % so that \hl and \st can wrap around \rf
\soulregister\cite{7} % so that \hl and \st can wrap around \cite
\renewcommand{\st}[1]{}

\usepackage{xr}

\usepackage{amsmath} % for mathematical symbols
\usepackage{stackengine} % for stacking elements

\newcommand\dbtilde[1]{\stackrel{\approx}{#1}}

\makeatletter
\newcommand*{\addFileDependency}[1]{% argument=file name and extension
  \typeout{(#1)}
  \@addtofilelist{#1}
  \IfFileExists{#1}{}{\typeout{No file #1.}}
}
\makeatother

\usepackage{textcomp} % for \textrightarrow
\usepackage{xifthen}
\usepackage{xcolor}
\usepackage{etoolbox}
\newboolean{togglechanges} 

\setboolean{togglechanges}{false}

\newcommand{\comment}[1]{\ifbool{togglechanges}
    {#1}  % updates-only version
    {\textcolor{blue}{#1}}}

\usepackage{bibentry}

\usepackage{graphicx}% Include figure files
\usepackage{dcolumn}% Align table columns on decimal point
\usepackage{bm}% bold math

\usepackage{textcomp} % for \textrightarrow
\usepackage{lipsum}

\begin{document}
\rmfamily

\title{Non-Abelian lattice gauge fields in the photonic synthetic frequency dimension}

\author{Dali Cheng$^{1,2}$, Kai Wang$^{3}$, Charles Roques-Carmes$^{1}$, Eran Lustig$^{1}$, Olivia Y. Long$^{1,4}$, Heming Wang$^{1}$, Shanhui Fan$^{1,2,4}$}
\email{shanhui@stanford.edu}

\affiliation{$^{1}$Edward L. Ginzton Laboratory, Stanford University, Stanford, California 94305, USA\looseness=-1}
\affiliation{$^{2}$Department of Electrical Engineering, Stanford University, Stanford, California 94305, USA\looseness=-1}
\affiliation{$^{3}$Department of Physics, McGill University, Montreal, QC H3A 2T8, Canada\looseness=-1}
\affiliation{$^{4}$Department of Applied Physics, Stanford University, Stanford, California 94305, USA\looseness=-1}

% \noindent	

% \noindent

\clearpage

%-----CHANGE SETUP FOR PARAGRAPH INDENTS AND SKIPS-----
% \setlength{\parindent}{0em}
% \setlength{\parskip}{.5em}
\vspace*{-2em}

%-------------------------------------
%------------- MAIN TEXT -------------
%-------------------------------------

%%%%%%%%%%%%%%%%%%%%%%%%%%%%%%%%%%%%%%%%%%%%%%%%%%%%%%%%%%%%%%%%%%%%%%%%%%%%%%%%%%%%%%%%%%%%%%%%%%%%%%%%%
%%% INSTRUCTIONS FOR COMMENTS AND MODIFICATIONS -- PLEASE RESPECT THEM, WILL SAVE A LOT OF TIME 
% - Make your comments either using Review tools or \comment{}
% - To refer to Figures: Fig.~\ref{fig:X}a
% - Same for Equations 
% - To Refer to SI sections, do manually for now 
%%%%%%%%%%%%%%%%%%%%%%%%%%%%%%%%%%%%%%%%%%%%%%%%%%%%%%%%%%%%%%%%%%%%%%%%%%%%%%%%%%%%%%%%%%%%%%%%%%%%%%%%%

\begin{abstract}
Non-Abelian gauge fields provide a conceptual framework for the description of particles having spins. The theoretical importance of non-Abelian gauge fields motivates their experimental synthesis and explorations. Here, we demonstrate non-Abelian lattice gauge fields for photons. In the study of gauge fields, lattice models are essential for the understanding of their implications in extended systems. We utilize the platform of synthetic frequency dimensions, which enables the study of lattice physics in a scalable and programmable way. We observe Dirac cones at time-reversal-invariant momenta as well as the direction reversal of eigenstate trajectories associated with such Dirac cones. Both of them are unique signatures of non-Abelian gauge fields in our lattice system. Our results highlight the implications of non-Abelian gauge field in the study of topological physics and suggest opportunities for the control of photon spins and pseudospins.
\end{abstract}

\maketitle

\section*{Introduction} 
Gauge fields are a foundational concept in physics, underlying a wide range of phenomena in electrodynamics, condensed matter physics, and particle physics~\cite{o2000gauge}. Non-Abelian gauge fields~\cite{yang1954conservation} have been of particular interest, since they interact with particles possessing internal spin degrees of freedom. Lattice models of gauge fields allow us to understand their physical implications in extended systems~\cite{kogut1979introduction}. 

The theoretical importance of gauge fields motivates their experimental synthesis and explorations. As of now, non-Abelian lattice gauge fields have been realized experimentally using cold atoms~\cite{wu2016realization, liang2024chiral} and electric circuits~\cite{wu2022non}. Yet, the demonstration of non-Abelian lattice gauge fields for \textit{photons} has not been achieved. Photons are fundamental particles for which artificial gauge fields can be synthetized~\cite{Fang2012, Rechtsman2013, Konstantin2015science, Bliokh2015, mittal2016measurement, ma2016spin, Chen2016, Zilberberg2018, miguel2018topological, lustig2019photonic, lumer2019light, chen2019non, yang2019synthesis, PhysRevLett.123.150503, dutt2020single, yang2020non, PhysRevA.103.013505, Guo2021, liu2021three, polimeno2021experimental, Leefmans2022, PhysRevLett.130.043803, cheng2023artificial, PhysRevLett.130.143801, Parto2023, zhang2023second, Jia2023, oliver2023sciadv, Yan:23, pang2024synthetic, yang2023non, PhysRevLett.132.143801, barsukova2024, Barczyk2024}. The demonstration of non-Abelian lattice gauge fields for photons could significantly enhance our understanding of the dynamical effects induced by them~\cite{yang2023non, Yan:23, yang2020non, liu2021three, pang2024synthetic, zhang2023second}. Such demonstration may also benefit photonic technologies by providing ways to control photon spins and pseudospins in topologically non-trivial ways~\cite{ozawa2019topological}.

Here, we demonstrate $\text{SU}(2)$ lattice gauge fields for photons in the platform of synthetic frequency dimensions~\cite{yuan2018synthetic, lustig2021topological}, an ideal playground to study lattice physics in a scalable and programmable way. We utilize a polarization-multiplexed fiber ring resonator, where polarization is taken as the pseudo-spin of photons. Non-Abelian physics is implemented via polarization rotations and polarization-dependent electro-optic modulations in the resonator. Key to our experiment is the theoretical observation that homogeneous non-Abelian lattice gauge potentials give rise to Dirac cones at time-reversal-invariant momenta in the Brillouin zone. Our experiments confirm the presence of non-Abelian lattice gauge fields by two experimental signatures: (1) the presence of such Dirac cones, and (2) the reversal of the directions of eigenstate trajectories along a loop in the Brillouin zone as the loop passes through the Dirac points. These results illustrate the interplay between non-Abelian gauge fields and topological physics.

\section*{Dirac cones induced by non-Abelian lattice gauge fields} 

\begin{figure*}
\centering
\vspace{-0.2cm}
  \includegraphics[scale=0.45]{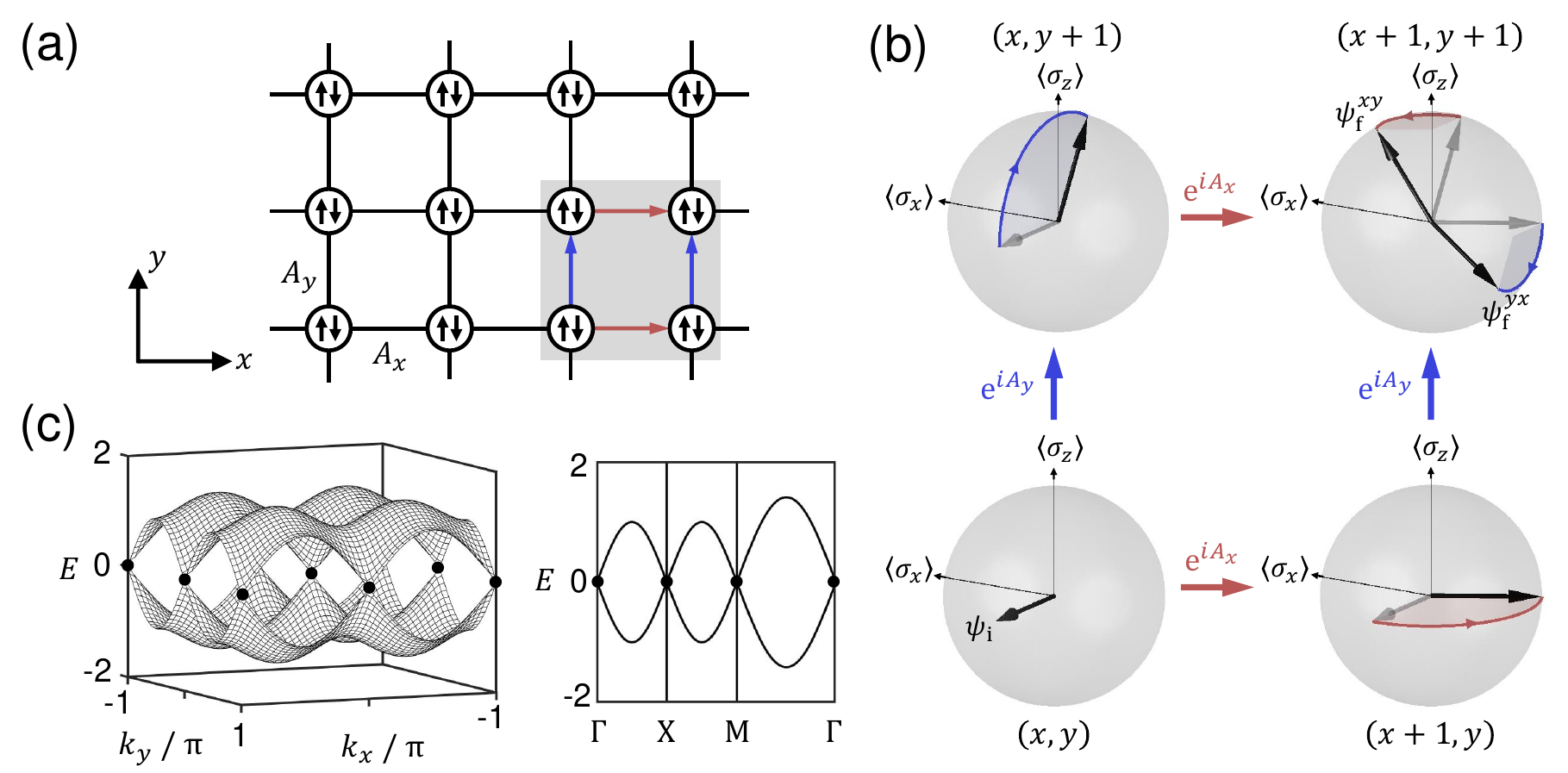}
    \caption{\small \textbf{Lattice model with non-Abelian gauge fields.} \textbf{(a)} Two-dimensional lattice model with non-Abelian gauge fields. Up- and down-arrows represent two orthogonal spin states on each lattice site. The square plaquette highlighted in grey is examined in \textbf{(b)}. \textbf{(b)} An illustration of the non-Abelian properties of the lattice gauge fields. A particle with initial spin state $\psi_i$ can take two alternative paths to hop from lattice site $(x,y)$ to $(x+1,y+1)$. The final spin state is dependent on the path because the spin rotations associated with the lattice links are not commutative. \textbf{(c)} Corresponding band structures of the lattice model in \textbf{(a)} with $A_x = (\pi/2) \sigma_z$ and $A_y = (\pi/2) \sigma_x$. $\Gamma:~(k_x,k_y)=(0,0)$, $X:~(k_x,k_y)=(\pi,0)$, $M:~(k_x,k_y)=(\pi,\pi)$. Dirac points at time-reversal-invariant momenta are highlighted by black dots.}
    \label{fig:1}
    \vspace{-0.3cm}
\end{figure*}

We consider the following lattice Hamiltonian, which describes the dynamics of a spin~\textonehalf~particle on a two-dimensional square lattice, as illustrated in Fig.~\ref{fig:1}(a):
\begin{equation}
    \hat{H} = \frac{1}{2} \sum_{x,y} \left( \hat{a}_{x+1,y}^\dagger \text{e}^{iA_x} \hat{a}_{x,y} + \hat{a}_{x,y+1}^\dagger \text{e}^{iA_y} \hat{a}_{x,y} + \text{H. c.} \right)
    \label{eq:hamiltonian-real},
\end{equation}
where $x, y\in\mathbb{Z}$ are lattice site indices, $\hat{a}_{x,y} = [\hat{a}_{x,y,\uparrow}, \hat{a}_{x,y,\downarrow}]^\text{T}$ is the particle's annihilation operator, and $A_x$ and $A_y$ are gauge potentials along the $x$ and $y$ directions, respectively. $A_x$ and $A_y$ are represented by 2$\times$2 Hermitian matrices, and the gauge fields are non-Abelian when $[A_x, A_y]\neq0$. $\text{e}^{iA_x}$ and $\text{e}^{iA_y}$ are link variables along the $x$ and $y$ directions, respectively.

The Hamiltonian of Eq.~(\ref{eq:hamiltonian-real}) describes path-dependent spin-rotations as a particle travels between lattice sites. Fig.~\ref{fig:1}(b) depicts an example as represented on the Bloch sphere, where $\sigma_{x(y,z)}$ is the Pauli matrix, and $\langle \sigma_{x(y,z)} \rangle$ axis corresponds to the $x (y, z)$ component of the spin state on the Bloch sphere. Consider the pathways through which a particle moves from the lattice site $(x,y)$ to the lattice site $(x+1, y+1)$. The particle may take a hop along the $x$ direction first, followed by a hop in the $y$ direction. Alternatively, the particle may take a pathway where the ordering of the hops is reversed. These two pathways yield final spin states of $\psi_f^{yx}=\text{e}^{iA_y}\text{e}^{iA_x}\psi_i$ and $\psi_f^{xy}=\text{e}^{iA_x}\text{e}^{iA_y}\psi_i$. These final spin states are different if $[A_x, A_y]\neq0$. Therefore, the lattice Hamiltonian of Eq.~(\ref{eq:hamiltonian-real}) exhibits non-Abelian gauge field physics originating from the lack of commutativity of gauge potentials.

The lattice Hamiltonian of Eq.~(\ref{eq:hamiltonian-real}) can be written in the reciprocal space as
\begin{equation}
    \hat{H}(k_x, k_y) = \cos(k_x\sigma_0 - A_x) + \cos(k_y\sigma_0 - A_y),
    \label{eq:hamiltonian-momentum}
\end{equation}
which is a Yang–Mills Hamiltonian~\cite{yang1954conservation, polimeno2021experimental, wu2022non}. $k_x$ and $k_y$ are the $x$ and $y$ components of the wavevector $\textbf{k}$, and $\sigma_0$ is the $2\times2$ identity matrix. Fig.~\ref{fig:1}(c) shows the band structure of this Hamiltonian with $A_x = (\pi/2) \sigma_z$ and $A_y = (\pi/2) \sigma_x$, where $\sigma_x$ and $\sigma_z$ are the Pauli matrices. 

The Hamiltonian of Eqs.~(\ref{eq:hamiltonian-real}) and (\ref{eq:hamiltonian-momentum}) exhibits two unique signatures associated with the non-Abelian nature of the lattice gauge fields. The first signature is the existence of Dirac points at $k_x,k_y\in\{0,\pi\}$. In our model of Eqs.~(\ref{eq:hamiltonian-real}) and (\ref{eq:hamiltonian-momentum}), the Dirac points at time-reversal-invariant momenta occur if and only if $[A_x,A_y]\ne0$ (with proof provided in the Methods section). The emergence of such Dirac points with linear band crossings is therefore a unique signature of the non-Abelian lattice gauge fields in our model. The second signature hinges on the topological charges carried by the Dirac points in the Hamiltonian of Eq.~(\ref{eq:hamiltonian-momentum}). This topological charge manifests itself in the eigenstate trajectory on the Bloch sphere as the wavevector $\textbf{k}$ varies along a closed loop in the two-dimensional Brillouin zone~\cite{bernevig2013topological, mark2014arxiv}. As detailed in the Methods section, when a wavevector loop passes the Dirac points, the handedness of the corresponding eigenstate trajectory reverses.

We make two observations about the theoretical results above: (1) Dirac points are common in two-dimensional systems~\cite{novoselov2005two, Geim2007, RevModPhys.81.109}. However, in many existing works Dirac points are observed only at time-reversal-breaking wavevectors, such as the \textit{K}- and \textit{K'}-points in a honeycomb lattice~\cite{novoselov2005two, Geim2007, RevModPhys.81.109}. For a photonic two-band model, having a Dirac point at time-reversal-invariant momenta, such as at the $\Gamma$ point of $\textbf{k} = (0,0)$, is unusual as it requires time-reversal-symmetry breaking. Note that the pseudo-spin of a photon is not changed under the time-reversal operation. While a linear dispersion can be synthesized at $\Gamma$ point in time-reversal-invariant systems, there always exist additional flat bands~\cite{Huang2011, PhysRevLett.114.223901, LiYang2021}, and such linear dispersions are accidental. The system of Eq.~(\ref{eq:hamiltonian-real}) exhibits a Dirac point at $\Gamma$ because the non-Abelian gauge fields break time-reversal symmetry. (2) A lattice system with a uniform Abelian gauge potential does not have any physically observable effects associated with the gauge potential, since the potential can always be made to vanish through gauge transformations. In contrast, the lattice system of Eq.~(\ref{eq:hamiltonian-real}) exhibits observable physical effects even though the non-Abelian gauge potentials are also uniform. These physical effects arise because the non-commutativity of the non-Abelian gauge potentials is gauge-invariant: it cannot be removed by any gauge transformation. The theoretical results above thus highlight the unique signatures that arise from the non-Abelian nature of the gauge potentials in the lattice system: these signatures cannot be achieved in an Abelian system.

\section*{Realization of $\text{SU}(2)$ lattice gauge fields in the photonic synthetic frequency dimension} 

\begin{figure}
\centering
\vspace{-0.2cm}
  \includegraphics[scale=0.45]{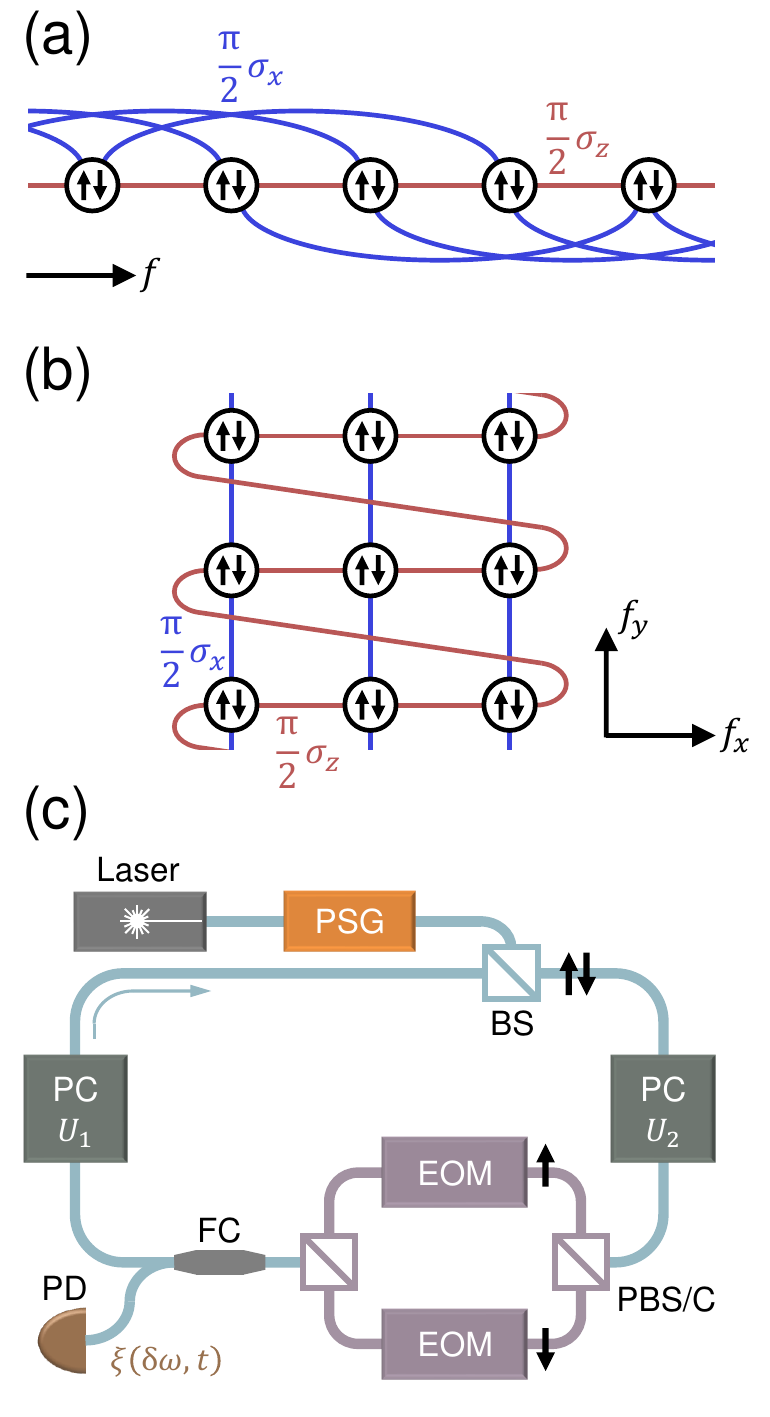}
    \caption{\small \textbf{Experimental setup to create non-Abelian lattice gauge fields in the photonic synthetic frequency dimension.} \textbf{(a)} The one-dimensional synthetic frequency lattice with $1$st- and $M$th-order couplings. $f$ represents the synthetic frequency axis. This lattice is generated when the resonator is subject to modulation signals containing both $\varOmega_R$ and $M\varOmega_R$ frequency harmonics. We use $M=3$ throughout the paper. \textbf{(b)} The corresponding two-dimensional synthetic frequency lattice with twisted boundary condition. $f_x$ and $f_y$ represent the two synthetic frequency axes. This lattice is equivalent to \textbf{(a)} and a direct implementation of Fig.~\ref{fig:1}(a). The non-Abelian gauge potentials $A_x = (\pi/2) \sigma_z$ and $A_y = (\pi/2) \sigma_x$ are labelled on the lattice links. \textbf{(c)} A schematic of the experimental setup. The up- and down-arrows represent the polarization (pseudo-spin) components supported by the fiber waveguide. The single-mode (polarization-maintaining) fiber waveguide is shown in blue (purple). PSG : polarization state generator, BS : beam splitter, PC : polarization controller, PBS/C : polarizing beam splitter/combiner, EOM : electro-optic modulator, FC : fiber coupler, PD : photodetector.}
    \label{fig:2}
    \vspace{-0.3cm}
\end{figure}

Motivated by the theoretical discussions in the previous section, here we seek to experimentally demonstrate non-Abelian lattice gauge fields with $A_x = (\pi/2) \sigma_z$ and $A_y = (\pi/2) \sigma_x$. For this purpose, a schematic of the experiment setup is shown in Fig.~\ref{fig:2}(c), with further details provided in the Methods section. Our experimental setup consists of a ring resonator made of optical fibers, where two orthogonal polarizations (horizontally and vertically polarized, represented in Fig.~\ref{fig:2} by $\uparrow$ and $\downarrow$, respectively) act as the pseudo-spin basis of the photons. The setup consists of three key parts: (1) an input laser and a polarization state generator for input state preparation; (2) two polarization controllers in the resonator implementing programmable SU(2) rotations $U_1$ and $U_2$ on the polarization state; and (3) polarization-maintaining branches where electro-optic modulation is performed on each of the two polarizations, separately. The resonator's output is sampled through a fiber coupler.

Figs.~\ref{fig:2}(a,b) illustrate the creation of a two-dimensional synthetic frequency lattice. The modulation signals in the resonator consist of both $\varOmega_R$ and $M\varOmega_R$ frequency tones, where $\varOmega_R$ is the free spectral range of the resonator. The $m$-th frequency mode in the resonator is thus coupled with $(m \pm 1)$-th modes via the modulation frequency component $\varOmega_R$, and with $(m \pm M)$-th modes via the frequency component $M\varOmega_R$. The resulting synthetic frequency lattice is shown in Fig.~\ref{fig:2}(a) with both short-range and long-range couplings. The one-dimensional lattice with long-range couplings in Fig.~\ref{fig:2}(a) is equivalent to a two-dimensional lattice with twisted boundary conditions as shown in Fig.~\ref{fig:2}(b)~\cite{yuan2018prb, Wang2020lsa}.

In our setup, the electro-optic modulators apply polarization-dependent transmission coefficients $\tau_{\uparrow,\downarrow}(t)$ on the two polarizations, with a modulation strength $g$ (details provided in the Methods section). We implement an interleaving modulation scheme, where the modulation signal consists of two concatenated sinusoidal waveforms of frequencies $\varOmega_R$ and $M\varOmega_R$, respectively, and the fundamental periodicity of $\tau_{\uparrow,\downarrow}(t)$ is $2T_R$, where $T_R=2\pi/\varOmega_R$ is the cavity roundtrip time. Within the two-dimensional synthetic frequency lattice, the non-Abelian gauge potentials are provided by static polarization rotations $U_1$ and $U_2$ in the resonator ($U_1=\text{e}^{-i\frac{\pi}{4}\sigma_y}$ and $U_2=\sigma_z$). In the weak modulation limit $g\ll\varOmega_R$, the evolution dynamics of the photon state in the resonator is governed by a Schr\"{o}dinger-like equation, with a Hamiltonian corresponding to Eq.~(\ref{eq:hamiltonian-real}) (see the Methods section for details of the derivation).  

In our experiments, we have $g = 0.14\Omega_R$ for band structure measurements and $g = 0.06 \Omega_R$ for eigenstate measurements. Both cases are outside the regime of validity of the weak modulation approximation. For a quantitative description of the system, influence of higher-order terms of $g$ need to be taken into account. Nevertheless, as shown in our simulations and experimental results below, the two key signatures associated with non-Abelian lattice gauge fields persist beyond the weak modulation approximation: the linear crossings at Dirac points in the band structure, and the reversal of the directions of eigenstate trajectories on a loop in the Brillouin zone as the loop passes through Dirac points. These results indicate the robustness of the non-Abelian lattice gauge fields outside the weak modulation regime.

\section*{Band structure measurements revealing the linear band crossings at Dirac points} 

\begin{figure*}
\centering
\vspace{-0.2cm}
  \includegraphics[scale=0.45]{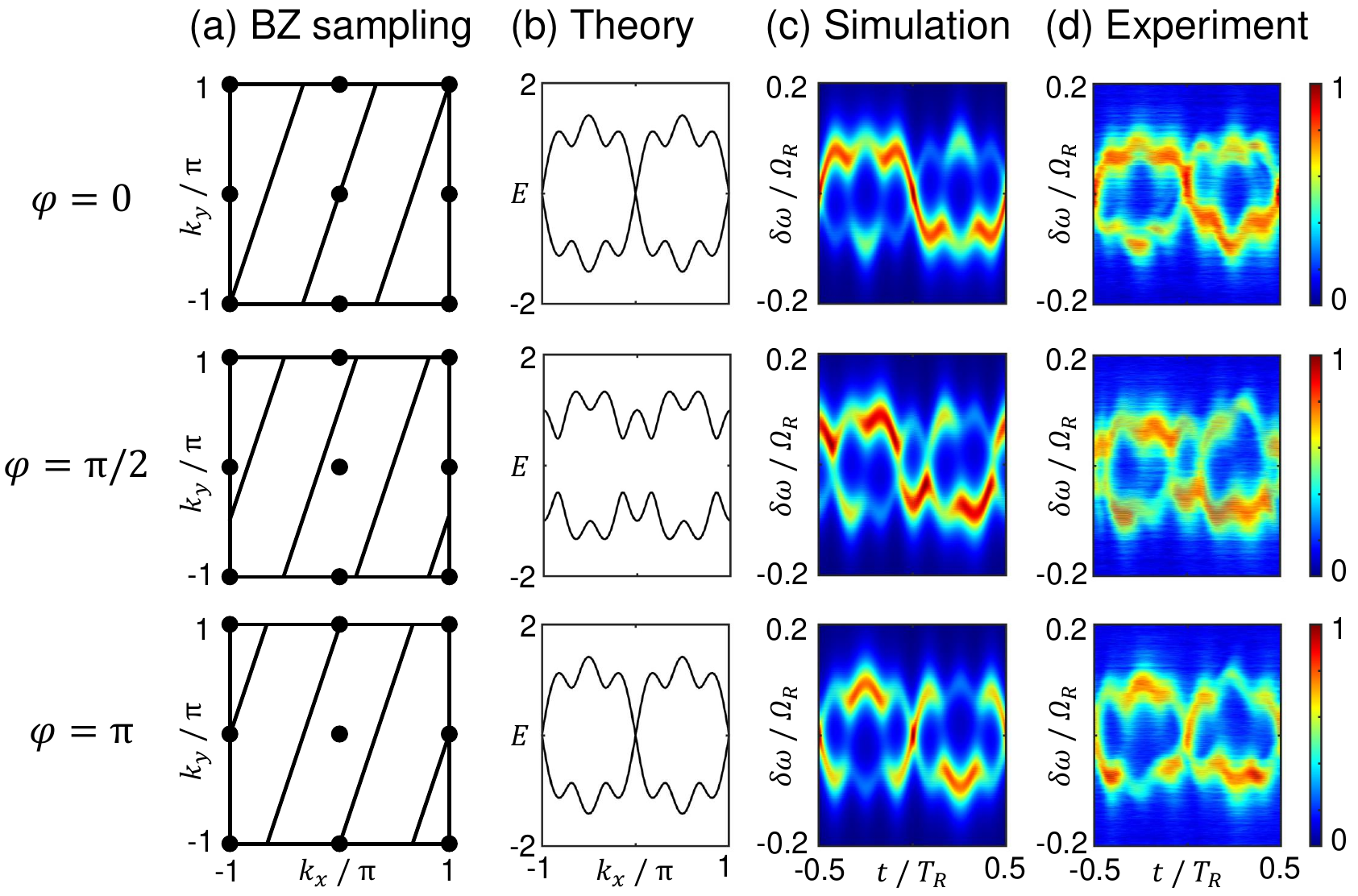}
    \caption{\small \textbf{Band structure measurements of the lattice Hamiltonian in the photonic synthetic frequency dimension.} \textbf{(a)} The sampling line in the two-dimensional Brillouin zone, with locations of Dirac points highlighted by black dots. \textbf{(b)} Theoretical band energies along the sampling line, calculated by diagonalizing the lattice Hamiltonian. \textbf{(c, d)} Simulated and measured results of time-dependent transmission spectrum of the system. In the simulation, we use the interleaving modulation scheme, take the modulation strength $g=0.14\varOmega_R$ and round trip power loss in the resonator as 1.74 dB. In both the simulation and experiment, the input state is fixed to be vertically polarized, and the color scale is normalized to $[0,1]$ for each individual subplot. The time origin in \textbf{(d)} is chosen such that the comparison with \textbf{(b)} and \textbf{(c)} is transparent.}
    \label{fig:3}
    \vspace{-0.3cm}
\end{figure*}

Band structure measurements are realized extending the methods in Refs.~\cite{dutt2019experimental, cheng2023multi}. Light intensity $\xi$ at the output photodetector is measured as a function of the input laser detuning $\delta\omega$ and time $t$. In a synthetic frequency lattice, $\delta\omega$ is interpreted as the Floquet quasi-energy of the system, and $t$ is interpreted as the wavevector $\textbf{k}$ since time is the reciprocal quantity of frequency~\cite{dutt2019experimental, cheng2023multi}. For a given time $t$, the resonant detuning $\delta\omega$ corresponds to the band energies at the wavevector $\textbf{k}$ associated with this particular time $t$. To realize band structure measurements in two-dimensional synthetic frequency lattices, we introduce a phase difference $\varphi$ between the two frequency tones $\varOmega_R$ and $M\varOmega_R$ in the modulation signal~\cite{cheng2023multi}. Band energies are then sampled along the one-dimensional line $L_\varphi: k_y = Mk_x + \varphi~(\text{mod }2\pi)$ in the two-dimensional Brillouin zone.

Fig.~\ref{fig:3} summarizes band structure measurement results with $\varphi\in \{0,\pi/2,\pi\}$. We first draw the sampling line in the two-dimensional Brillouin zone in Fig.~\ref{fig:3}(a). Trajectories parametrized by $\varphi\in \{0,\pi\}$ intersect the $\Gamma$ point with $k_x = k_y = 0$ and the $M$ point with $k_x = k_y = \pi$ (denoted by black dots). The corresponding theoretical band energies along the sampling line are shown in the top and bottom panels of Fig.~\ref{fig:3}(b), where the linear band crossings are evident at both $\Gamma$ and $M$ points. For $\varphi=\pi/2$, the sampling line avoids the $\Gamma$ and the $M$ points, and therefore the band structure is gapped (middle panel, Fig.~\ref{fig:3}(b)). We then show in Figs.~\ref{fig:3}(c,d) simulations and experimental measurements of the time-dependent transmission spectra $\xi(\delta\omega,t)$ of the system (more details provided in the Methods section), for a fixed input polarization state of light. In the simulations we do not assume the weak modulation approximation. Nevertheless, the features of linear band crossings at $\Gamma$ and $M$ points are well preserved. And our experimental measurements display resonant features in the spectra that closely match the theoretical band structures and simulations. Dirac band crossings and band gaps are clearly observed in our experiment for $\varphi \in \{ 0,\pi \}$ and $\varphi=\pi/2$, respectively. The non-uniformity of the signal strength at resonance originates from the variation in the overlap between the eigenstate and the fixed input polarization. This observation hints at a projective method for eigenstate measurements by exciting the system with different polarization inputs, which we describe and utilize in the following section.

\section*{Tomographic eigenstate reconstructions revealing reversed directions of eigenstate trajectories}

To reveal the eigenstate trajectory in our lattice model, we perform eigenstate tomographic reconstructions. The steady-state polarization in the resonator at a given wavevector $\textbf{k}$, represented by a Jones vector $\psi_\text{ss}$, is related to the input polarization state $\psi_\text{in}$ by the Green’s function of the system:
\begin{equation}
    \psi_\text{ss}(\delta\omega, \textbf{k}, \psi_\text{in}) = \frac{1}{\delta\omega - \hat{H}(\textbf{k}) + i\gamma_0/2} \psi_\text{in},
\end{equation}
where $\gamma_0$ is the intrinsic loss rate in the resonator. By expanding $\hat{H}(\textbf{k}) = \omega_{-}(\textbf{k}) \psi_{-}(\textbf{k}) \psi_{-}^{\dagger}(\textbf{k}) ~+~ \omega_{+}(\textbf{k}) \psi_{+}(\textbf{k}) \psi_{+}^{\dagger}(\textbf{k})$, where $\omega_{\pm}(\textbf{k})$ is the eigenenergy of the upper/lower band and $\psi_{\pm}(\textbf{k})$ is the associated eigenstate, and taking the resonant condition $\delta\omega = \omega_{-}(\textbf{k})$, we get $\xi[\delta\omega=\omega_{-}(\textbf{k}),\textbf{k},\psi_\text{in}] \approx 4\gamma_0^{-2} |\psi_{-}^{\dagger}(\textbf{k}) \psi_\text{in}|^2$.
This approximation is valid when $\gamma_0 \ll |\omega_{+}(\textbf{k}) - \omega_{-}(\textbf{k})|$, and thus the steady state is dominated by the eigenstate $\psi_{-}(\textbf{k})$. Therefore, by extracting the on-resonance intensities in the time-dependent transmission spectra $\xi(\delta\omega, \textbf{k}, \psi_\text{in})$ for different input polarizations $\psi_\text{in}$, we can reconstruct the Stokes parameters of the eigenstate. For instance, the first Stokes parameter $\langle\sigma_x\rangle(\textbf{k})$ of the eigenstate $\psi_{-}(\textbf{k})$ can be obtained by subtracting two projective measurements: $\langle\sigma_x\rangle (\textbf{k}) = (\gamma_0^2/4) [ \xi_X(\textbf{k})- \xi_Y(\textbf{k})]$, where $\xi_P(\textbf{k}) = \xi[\delta\omega=\omega_{-}(\textbf{k}),\textbf{k},\psi_\text{in}=P]$, $P\in\{H$ (horizontal), $V$ (vertical), $X$ ($+45^{\circ}$ linear polarized), $Y$ ($+135^{\circ}$ linear polarized), $L$ (left circular polarized), $R$ (right circular polarized)$\}$.

\begin{figure*}
\centering
\vspace{-0.2cm}
  \includegraphics[scale=0.45]{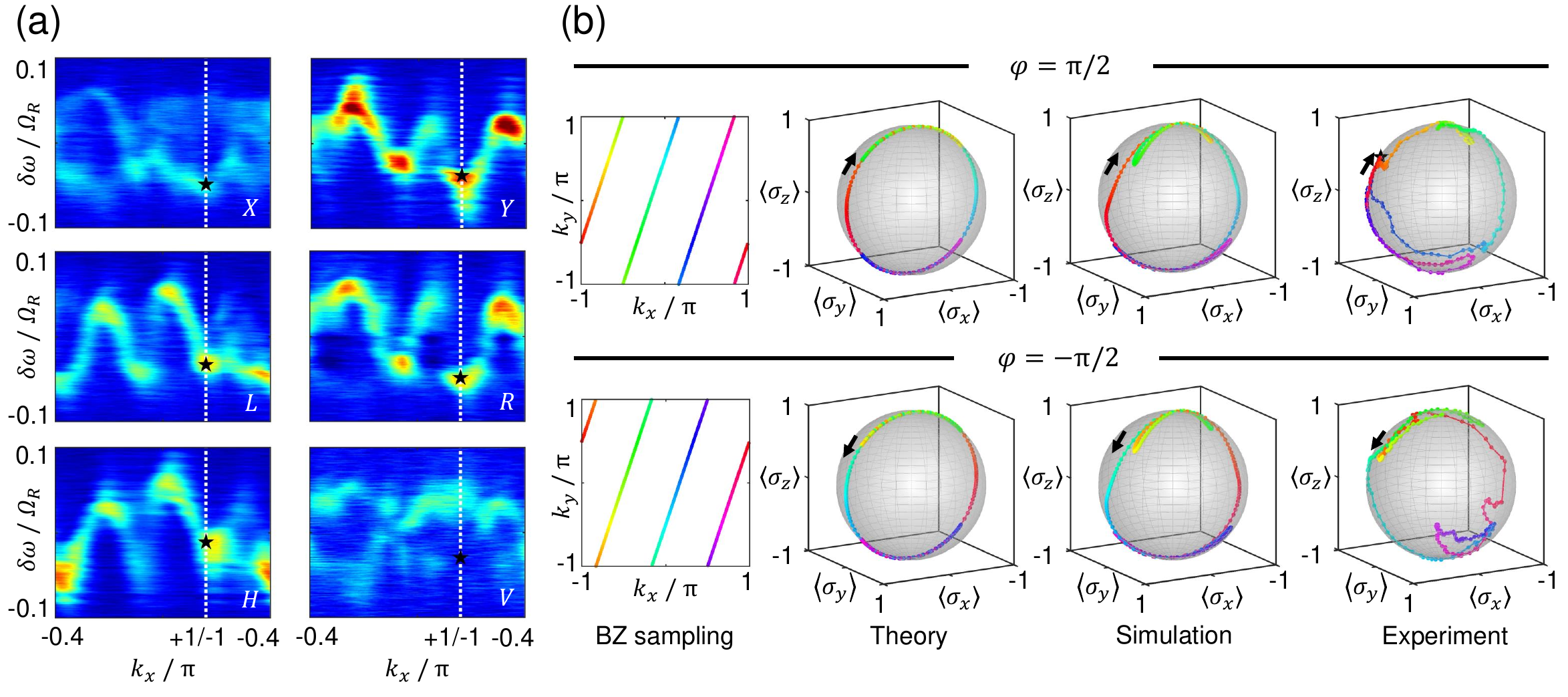}
    \caption{\small \textbf{Tomographic eigenstate reconstructions and observation of eigenstate trajectories.} \textbf{(a)} Measured wavevector-dependent transmission spectra $\xi(\delta\omega,\textbf{k},\psi_\text{in})$ with different input polarization states $\psi_\text{in}$, for $\varphi=\pi/2$. The input polarization state $\psi_\text{in} \in \{X,Y,L,R,H,V\}$ is indicated at the bottom-right corner of each plot. The white dashed line is drawn at $k_x = \pm \pi$, and the black stars coincide with maximum values of $\xi$ along the white dashed line in the region $\delta\omega<0$, i.e., the lower band. All six plots share the same color scale. \textbf{(b)} Reconstructed eigenstate trajectories along the loops $L_{\varphi=\pm\pi/2}$ for the lower band. An eigenstate on the Bloch sphere is related to its location in the two-dimensional Brillouin zone on the left-most panel by its color. Theoretical results are obtained by diagonalizing the lattice Hamiltonian. Simulation (experiment) results are obtained by eigenstate tomographic reconstructions on simulated (experimental) data of $\xi(\delta\omega,\textbf{k},\psi_\text{in})$. Note that the same global SO(3) rotation is applied to the experimental trajectories of $\varphi=\pi/2$ and $\varphi=-\pi/2$ to facilitate its comparison with theoretical and simulation results. This global rotation originates from the difference between the polarization basis definitions in the PSG and in the resonator. In simulations we take $g=0.06\varOmega_R$, and the round trip power loss in the resonator as 0.87 dB. The wavevector (time) origin of the experimental data is chosen to match that of the eigenstate trajectories in theory and simulation.}
    \label{fig:4}
    \vspace{-0.3cm}
\end{figure*}

Fig.~\ref{fig:4}(a) shows projective measurements of time-dependent transmission spectra $\xi(\delta\omega,\textbf{k},\psi_\text{in})$ with six different input polarizations ${\psi_\text{in}}$ and $\varphi=\pi/2$. By extracting the on-resonance values of $\xi(\delta\omega,\textbf{k},\psi_\text{in})$, we are able to reconstruct the normalized Stokes vector $(\langle\sigma_x\rangle,\langle\sigma_y\rangle,\langle\sigma_z\rangle)(\textbf{k})$ of the eigenstate of the lower band. In Fig.~\ref{fig:4}(b), in the first column we show the loops formed by the sampling line $L_{\varphi=\pm\pi/2}$ in the two-dimensional Brillouin zone. In the second column, we show the eigenstate trajectories of the lower band of Hamiltonian Eq.~(\ref{eq:hamiltonian-momentum}) along the loops $L_{\varphi=\pm\pi/2}$. Comparing the eigenstate trajectories of $\varphi=\pi/2$ and $\varphi=-\pi/2$, we find that their winding directions are opposite, which is a signature of the non-Abelian lattice gauge fields in our model (see the Methods section for more discussions). The third column shows simulations of the eigenstate trajectories (details described in the Methods section), where the weak modulation approximation on $g$ is not applied. The signature, that the eigenstate trajectories on the Bloch sphere have reversed directions for $\varphi=\pi/2$ and $\varphi=-\pi/2$, persists outside the weak modulation approximation. The last column in Fig.~\ref{fig:4}(b) shows experimental results of the eigenstate trajectories along loops $L_{\varphi=\pm\pi/2}$, and we observe the expected reversal of winding directions. These results further confirm the emergence of non-Abelian lattice gauge fields for photons in the synthetic frequency dimension.

\section*{Discussion}

We experimentally observe non-Abelian lattice gauge fields for photons using synthetic frequency dimensions created in a photonic resonator system. We combine polarization rotations and polarization-dependent modulations to realize the non-Abelian nature of the lattice gauge fields. Our system realizes a two-dimensional Yang–Mills Hamiltonian. Within this model, the non-Abelian lattice gauge fields exhibit two unique signatures that we observe experimentally: the emergence of Dirac points with linear band crossings at time-reversal-invariant momenta in the Brillouin zone, and the reversal of the directions of the eigenstate trajectories on a loop in the Brillouin zone as the loop passes through the Dirac points. We emphasize that the method we developed in this paper allows us to perform the first fully tomographic reconstruction of eigenstates in a synthetic frequency lattice. Compared to existing eigenstate measurement techniques~\cite{li2023direct, pellerin2023wavefunction}, our method does not require prior information on the structure of eigenstates from the model, and therefore is generalizable to other types of lattice Hamiltonians. Our method is facilitated by the usage of polarization as the pseudo-spin degree of freedom of photons and could take advantage of other polarization measurement schemes. 

The photonic synthetic frequency dimension is an ideal platform to observe and engineer non-Abelian lattice physics, thanks to their programmability and the enormous bandwidth of photonic systems. Large-scale photonic lattices with more than $10^5$ lattice sites can now be implemented~\cite{senanian2023programmable}. In addition to the lattice model in this paper, our setup could be extended to capture the physics of more complex Hamiltonians, for example, those possessing non-Abelian scalar~\cite{chen2019non} and non-Abelian non-Hermitian gauge potentials~\cite{pang2024synthetic}. Band structure measurements could be used to reveal more topological features related to non-Abelian lattice gauge fields such as spin–orbit-coupled butterfly pairs~\cite{yang2020non}, trios~\cite{liu2021three}, and the Hofstadter moth~\cite{osterloh2005cold}. By monitoring the evolution dynamics of wavepackets in the photonic synthetic frequency lattice, one might be able to also experimentally observe the Zitterbewegung effect~\cite{zhang2008observing, vaishnav2008observing} and the non-Abelian Aharonov–Bohm effect~\cite{horvathy1986non}. Our results open up new possibilities in studying non-Abelian lattice gauge physics, their dynamics, and their connection with topological phases of matter.

\section{Authors contributions}
D. C., K. W., and S. F. conceived the original idea. D. C., K. W., and E. L. designed the experiment. D. C. and C. R.-C. acquired the data with contributions from O. Y. L.. D. C. and C. R.-C. analyzed the data. D. C. performed theoretical and numerical studies with contributions from K. W. and H. W.. S. F. supervised the research. D. C., C. R.-C., and S. F. wrote the manuscript with inputs from all authors.

\section{Competing interests}
The authors declare no potential competing financial interests.

\section{Data and code availability statement}
The data and codes that support the plots within this paper and other findings of this study are available from the corresponding author upon reasonable request.

\section{Acknowledgements}
This work is supported by MURI projects from the U.S. Air Force Office of Scientific Research (Grants No. FA9550-18-1-0379 and FA9550-22-1-0339). We thank Professor David A. B. Miller for providing laboratory space and equipment. The authors would like to acknowledge discussions with Haiwen Wang, Professor Yi Yang, and Professor Kasper Van Gasse, and help from Lawton Skaling with experiments. K. W. acknowledges financial support from Québec’s Ministère de l'Économie, de l'Innovation et de l'Énergie. C.~R.-C. is supported by a Stanford Science Fellowship.

\bibliography{bibliography}
\bibliographystyle{naturemag}

\newpage
\onecolumngrid

\section{Extended Figures}

\begin{figure}[h]
\centering
\vspace{-0.2cm}
  \includegraphics[scale=0.4]{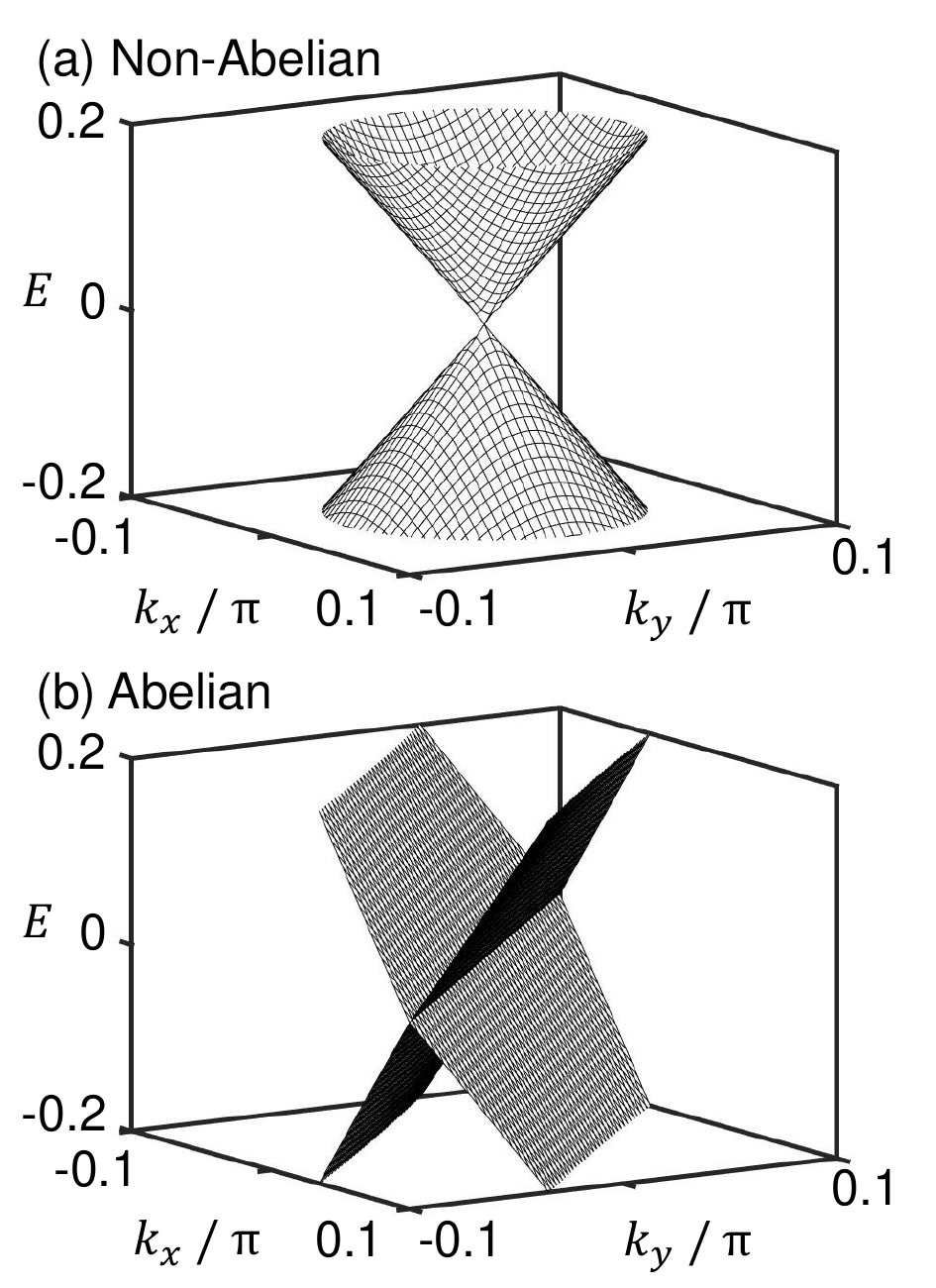}
    \caption{\small \textbf{Dirac cone as a signature of non-Abelian lattice gauge fields.} Theoretical band structure of Hamiltonian Eq.~(\ref{eq:hamiltonian-momentum}) in the vicinity of $\Gamma:(k_x,k_y)=(0,0)$, for \textbf{(a)} non-Abelian gauge potentials $A_x = (\pi/2) \sigma_z$ and $A_y = (\pi/2) \sigma_x$, and \textbf{(b)} Abelian gauge potentials $A_x=A_y=(\pi/2) \sigma_z$.}
    \label{fig:S1}
    \vspace{-0.3cm}
\end{figure}
\clearpage

\newpage
\begin{figure}
\centering
\vspace{-0.2cm}
  \includegraphics[scale=0.5]{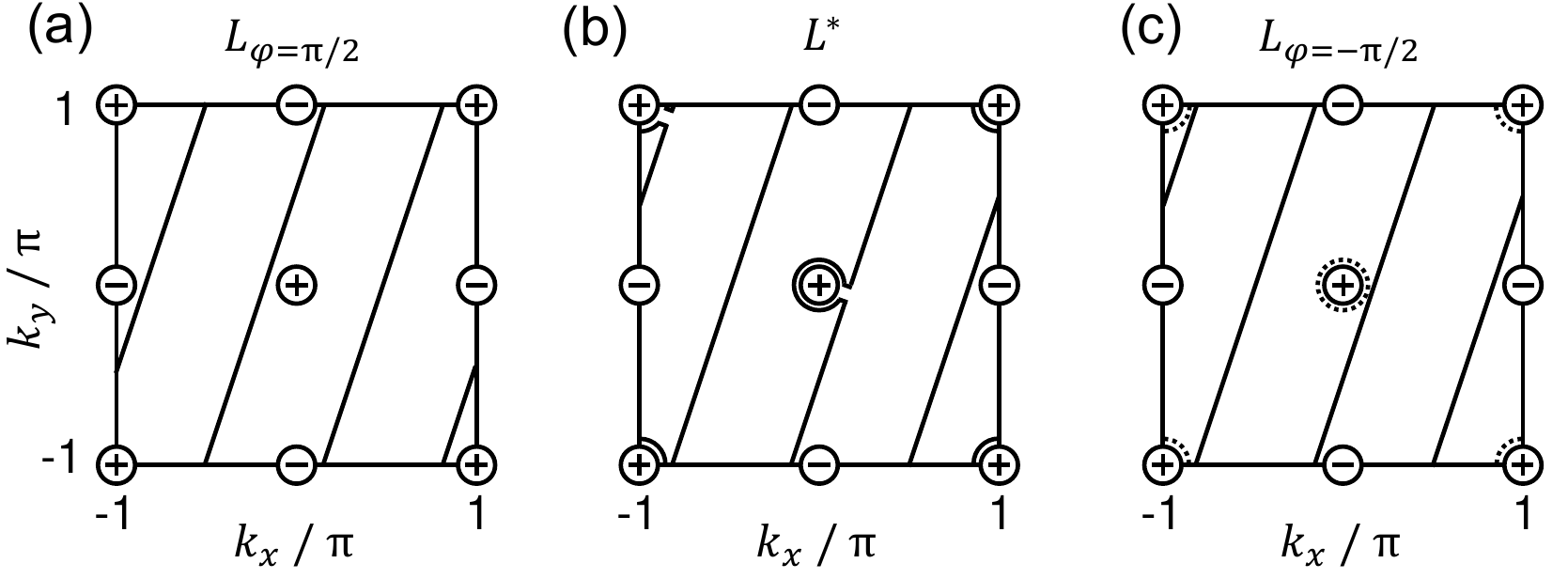}
    \caption{\small \textbf{Illustration of loops in the Brillouin zone and topological charges of Dirac points.} The Dirac points at $\Gamma:(0,0)$, $M:(\pi,\pi)$ and those at $X:(\pi,0), (0,\pi)$ are of opposite charges, as indicated by the $\pm$ signs. \textbf{(a)} The black solid line represents the sampling loop $L_{\varphi=\pi/2}$. \textbf{(b)} An auxiliary loop $L^{*}$ by moving $L_{\varphi=\pi/2}$ to the $+k_x$ direction by $\pi/3$ but without touching any Dirac points. The loop hence circles around Dirac points at $\Gamma$ and M. The eigenstate trajectories have the same handedness for $L_{\varphi=\pi/2}$ and $L^{*}$. \textbf{(c)} The black solid line represents the sampling loop $L_{\varphi=-\pi/2}$. Note that $L_{\varphi=-\pi/2}$ can be viewed as the remaining part of $L^{*}$ after the circles in the vicinity of Dirac points (as shown in dashed lines) are disconnected. Thus the difference between the eigenstate trajectories along $L_{\varphi=\pm\pi/2}$ exists in the topological charges of Dirac points at $\Gamma$ and M.}
    \label{fig:S5}
    \vspace{-0.3cm}
\end{figure}
\clearpage

\newpage
\begin{figure}
\centering
\vspace{-0.2cm}
  \includegraphics[scale=1.0]{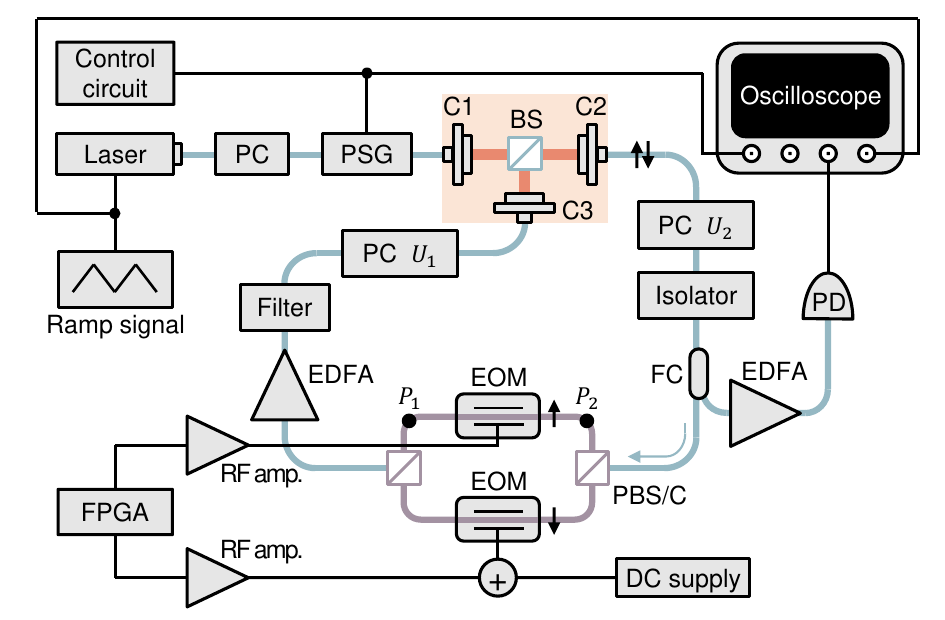}
    \caption{\small \textbf{A detailed illustration of the experimental setup.} Electronic connections are shown in black thin lines, single-mode fibers in blue lines, polarization-maintaining fibers in purple lines, and free-space collimated beams in orange thick lines. The free-space section is highlighted in the orange box. PC: polarization controller, PSG: polarization state generator, C1–C3: fiber–free-space collimators, BS: beam splitter, FC: fiber coupler, PBS/C: polarizing beam splitter/combiner, EOM: electro-optic modulator, EDFA: erbium-doped fiber amplifier, FPGA: field-programmable gate array, RF amp.: radiofrequency amplifier, PD: photodetector.}
    \label{fig:S2}
    \vspace{-0.3cm}
\end{figure}
\clearpage

\newpage
\begin{figure}
\centering
\vspace{-0.2cm}
  \includegraphics[scale=0.4]{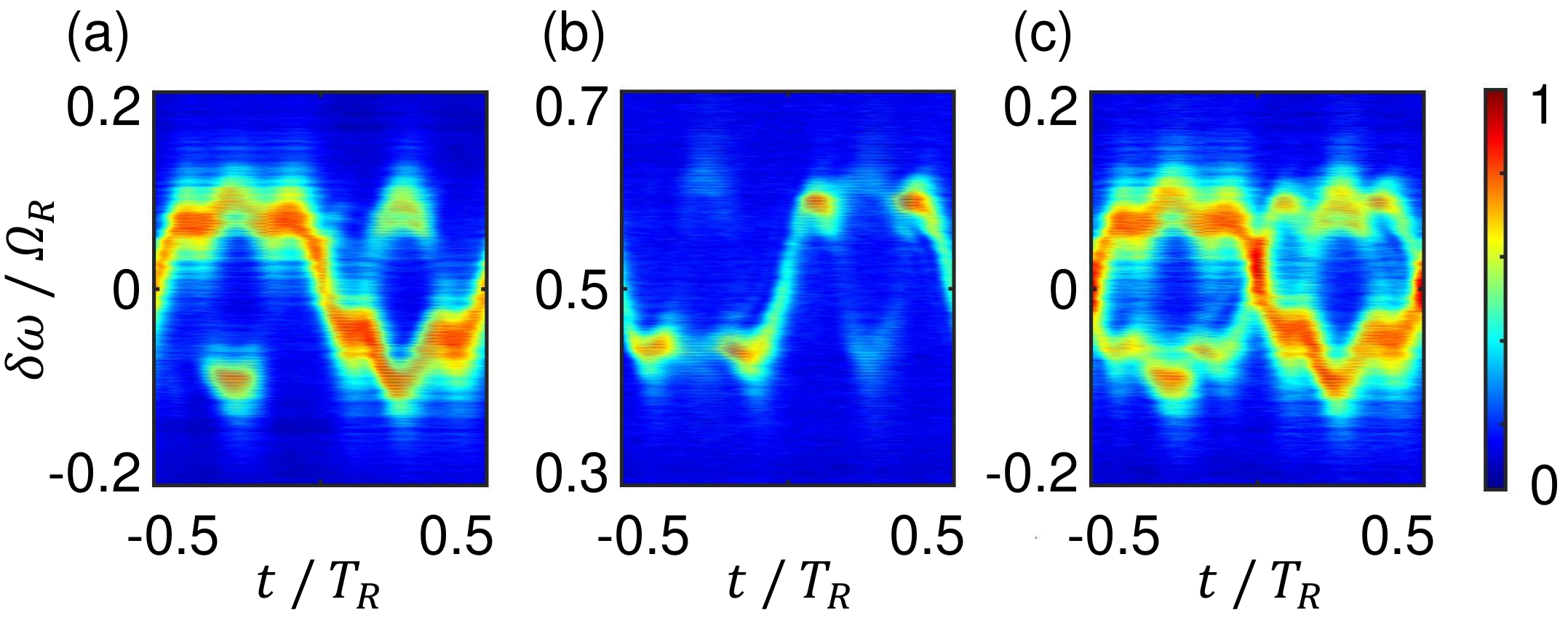}
    \caption{\small \textbf{Comparison between spectra $\Tilde{\xi}(\delta\omega, t)$ and $\dbtilde{\xi}(\delta\omega, t)$.} The definitions of $\Tilde{\xi}$ and $\dbtilde{\xi}$ can be found in the Methods section. \textbf{(a)} The resonance features in $\Tilde{\xi}$ near $\delta\omega=0$. \textbf{(b)} The resonance features in $\Tilde{\xi}$ near $\delta\omega=0.5\Omega_R$. \textbf{(c)} The quantity $\dbtilde{\xi}$ as defined in Eq.~(\ref{eq:double-xi-1}) in the Methods section. \textbf{(a)} and \textbf{(b)} are experimentally obtained with $\varphi=0$, $g=0.14\Omega_R$, and vertically polarized input light. \textbf{(c)} is obtained by averaging \textbf{(a)} and \textbf{(b)}, and is a replica of Fig. 3(d). The color scale is normalized to $[0,1]$ for each individual subplot.}
    \label{fig:S3}
    \vspace{-0.3cm}
\end{figure}
\clearpage

\newpage
\begin{figure}
\centering
\vspace{-0.2cm}
  \includegraphics[scale=0.4]{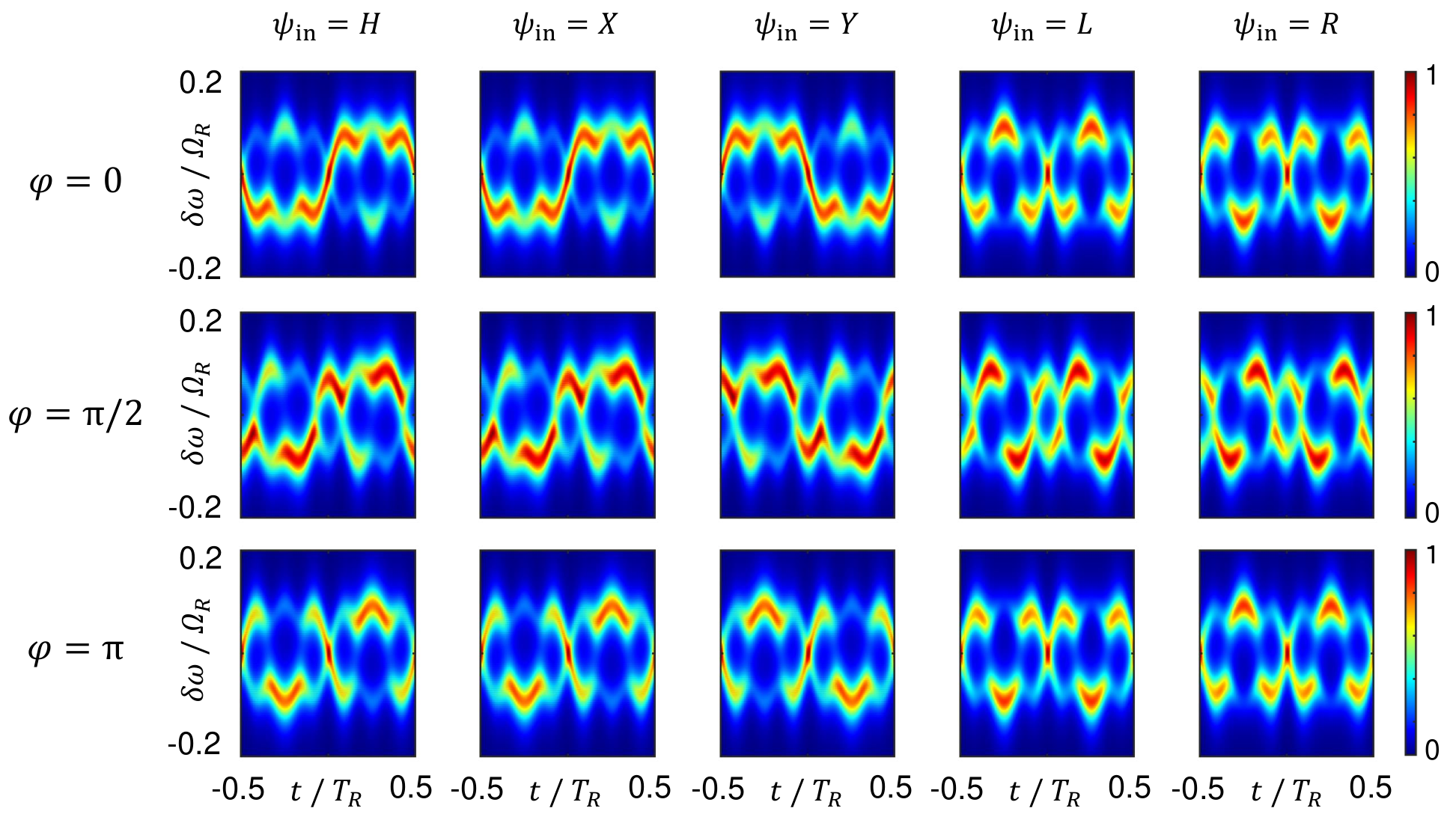}
    \caption{\small \textbf{Simulations of band structure measurements with different input polarizations.} We calculate and plot $\dbtilde{\xi}(\delta\omega, t)$ as described in this figure. The input polarizations are set to be $H$ (horizontal), $X$ ($+45^{\circ}$ linear polarized), $Y$ ($+135^{\circ}$ linear polarized), $L$ (left circular polarized), and $R$ (right circular polarized). The results of vertically polarized input are shown in Fig. 3(c). We take the modulation strength $g=0.14\varOmega_R$ and round trip power loss in the resonator as 1.74 dB. For each value of $\varphi$, the color scale is the same as that in Fig. 3(c) to facilitate the comparison between conditions where $\varphi$ is identical but $\psi_\text{in}$ is different.}
    \label{fig:S4}
    \vspace{-0.3cm}
\end{figure}
\clearpage

\section{Methods}

\section*{Signatures of non-Abelian lattice gauge fields} 
\label{sec:theory}

In this section, we theoretically discuss the two key signatures of interest of the non-Abelian lattice gauge fields by analyzing the Hamiltonian Eq.~(\ref{eq:hamiltonian-momentum}). These experimentally observable signatures can be captured by the lattice model of Eq.~(\ref{eq:hamiltonian-momentum}).

\textbf{Dirac cones with linear band crossings at time-reversal-invariant momenta in the Brillouin zone.} We prove that in the lattice model Eq.~(\ref{eq:hamiltonian-momentum}), the Dirac cones at time-reversal-invariant momenta occur if and only if the gauge potentials are non-Abelian. Therefore, the linear band crossings at $\Gamma$, $X$, and $M$ points are unique signatures of the non-Abelian lattice gauge fields. As an illustration of this result, Fig.~\ref{fig:S1} compares band dispersions near the $\Gamma$ point for two different cases: in Fig.~\ref{fig:S1}(a), the gauge potential is non-Abelian and the Dirac cone emerges; in Fig.~\ref{fig:S1}(b), the gauge potential is Abelian and the Dirac cone structure is absent. In the following, we provide a formal proof. In reciprocal space, the Hamiltonian of interest is
\begin{equation}
    \hat{H}(k_x, k_y) = \cos(k_x\sigma_0 - A_x) + \cos(k_y\sigma_0 - A_y).
    \label{eq:S1}
\end{equation}
Since the Hamiltonian is Hermitian, it can be parametrized without loss of generality by
\begin{equation}
    A_x = \frac{\pi}{2} \textbf{d}_x \cdot \boldsymbol{\sigma},
    A_y = \frac{\pi}{2} \textbf{d}_y \cdot \boldsymbol{\sigma},
    \label{eq:S2}
\end{equation}
where $\textbf{d}_{x},\textbf{d}_{y}\in\mathbb{S}^2$ are real vectors of unit length. The vectors $\textbf{d}_{x},\textbf{d}_{y}$ are the axes of spin rotations associated with the link variables $\text{e}^{iA_x}, \text{e}^{iA_y}$. To establish our result, it is sufficient to prove that
(1) $[A_x, A_y] \ne 0~\Leftrightarrow~\textbf{d}_x \neq \pm \textbf{d}_y$; 
(2) Emergence of Dirac points at time-reversal-invariant momenta $\Leftrightarrow~\textbf{d}_x \neq \pm \textbf{d}_y$. The proof of (1) is straightforward by calculating the commutator of the gauge potentials:
\begin{equation}
    [A_x, A_y] = i\frac{\pi^2}{2} (\textbf{d}_x \times \textbf{d}_y) \cdot \boldsymbol{\sigma},
    \label{eq:S3}
\end{equation}
which is non-zero if and only if $\textbf{d}_x \neq \pm \textbf{d}_y$.

To prove (2), recall that Dirac points feature ellipse-shaped iso-energy contours around them. In the vicinity of a time-reversal-invariant momentum, for example, $k_x=k_y=0$, we expand the Hamiltonian Eq.~(\ref{eq:S1}) to the leading order of $k_x$ and $k_y$:
\begin{equation}
    \hat{H}(k_x, k_y) = k_x \sin(A_x) + k_y \sin(A_y).
    \label{eq:S4-sin}
\end{equation}
According to Eq.~(\ref{eq:S2}), $A_x$ has eigenvalues $\pm\pi/2$, and we can write $A_x = V
\begin{bmatrix} \pi/2 & 0\\ 0 & -\pi/2 \end{bmatrix}
V^{\dagger}$ where $V$ is the unitary matrix that diagnolizes $A_x$. Hence, $\sin(A_x) = V
\begin{bmatrix} \sin(\pi/2) & 0\\ 0 & \sin(-\pi/2) \end{bmatrix}
V^{\dagger} = (2/\pi) A_x$. Similarly, we have $\sin(A_y) = (2/\pi) A_y$. The Hamiltonian therefore can be written as
\begin{equation}
    \hat{H}(k_x, k_y) = \frac{2}{\pi}(k_x A_x + k_y A_y).
    \label{eq:S4}
\end{equation}
By diagonalizing Eq.~(\ref{eq:S4}), we find the iso-energy contours in the form
\begin{equation}
    k_x^2 + k_y^2 +
    2(\textbf{d}_x\cdot\textbf{d}_y)k_xk_y = \text{const.},
    \label{eq:S5}
\end{equation}
which represents an ellipse if and only if $\textbf{d}_x \neq \pm \textbf{d}_y$. This completes our proof.

\textbf{Reversed directions of eigenstate trajectories associated with topological charges of Dirac points.} The Hamiltonian Eq.~(\ref{eq:S1}) has the chiral symmetry $\sigma_y \hat{H}(k_x, k_y) \sigma_y^{-1} = -\hat{H}(k_x, k_y)$. Therefore the eigenstates are pinned onto a circle defined as the intersection of the $\langle \sigma_x \rangle$–$\langle \sigma_z \rangle$ plane with the Bloch sphere. When traced along a closed loop in the Brillouin zone, an eigenstate thus traces out a trajectory that winds around the origin of the Bloch sphere for an integer number of times. If such a closed loop in the Brillouin zone encircles a Dirac point, the winding of the trajectory can be associated with the topological charge of the Dirac point~\cite{bernevig2013topological, mark2014arxiv}. An eigenstate trajectory completing a single winding in the counter-clockwise (clockwise) direction indicates that the Dirac point enclosed in the interior of the loop has topological charge $+1$ ($-1$). The sampling line of our interest, $L_{\varphi}: k_y = Mk_x + \varphi~(\text{mod }2\pi)$, forms a non-contractible loop that cannot be characterized by an ``interior'' and an ``exterior.'' In this case, the Dirac point is manifested by the change in the eigenstate trajectory as the sampling line moves across the Dirac point, as illustrated in Fig.~\ref{fig:S5}. In Hamiltonian Eq.~(\ref{eq:S1}), the Dirac points at $\Gamma$, $M$ and those at $X$ are of opposite charges. The loop $L_{\varphi=\pi/2}$ crosses two Dirac points of the same charge as it is moved to $L_{\varphi=-\pi/2}$. As a result, the handedness of the eigenstate trajectory changes from clockwise to counter-clockwise. This corresponds to an absolute value of topological charge of 2, the sum of the charges of the Dirac points at $\Gamma$ and at $M$.

\section*{Experimental setup} 
\label{sec:expt}
In this section, we provide further details on the experimental setup to observe non-Abelian lattice gauge fields in the photonic synthetic frequency dimension.

Fig.~\ref{fig:S2} is a detailed illustration of our experimental setup. The ring resonator in our setup is based on optical fibers, with a free spectral range of $\varOmega_R=2\pi\times4~\text{MHz}$. The continous-wave laser (Grade 3 Orion laser from Redfern Intergrated Optics) has a center wavelength of $1542.057~\text{nm}$ and a linewidth of $2.8~\text{kHz}$. The laser frequency is swept in a range of approximately $20\varOmega_R$ by a $600~\text{mVpp}$, $50~\text{Hz}$ ramp signal from a function generator. The polarization state generator (from Luna Inc) is programmed by an Arduino microcontroller through driving chips from Texas Instruments, and produces specific input polarization states for the resonator. The input–output coupling of the resonator is implemented using a free-space section, as highlighted by the orange box in Fig.~\ref{fig:S2}. The free-space collimated beams are coupled to single mode fibers via aspherical lenses. The input light beam is coupled into the resonator via a non-polarizing $90:10$ beam splitter (reflection coefficient: $90\%$, transmission coefficient: $10\%$). In the resonator light propagates in the clockwise direction, and an isolator is used to remove the counter-clockwise-propagating mode. The main section of the resonator are single-mode fibers supporting both polarizations. We employ two polarization controllers $U_1$ and $U_2$ (from Agilent, motorized and programmable) in the main section to produce SU(2) rotations on the polarization state of light. 

We also incorporate a polarization-dependent modulation section in the resonator. Two orthogonal polarizations in the main section are split via a fiber-coupled polarizing beam splitter into two branches of polarization-maintaining fibers. We refer to these two branches as "spin-up" and "spin-down" branches below. Within each branch we insert an electro-optic phase modulator based on lithium niobate waveguides (from Optilab, $5~\text{GHz}$ bandwidth). The modulation signals are generated by an FPGA (Moku:Lab from Liquid Instruments) and then amplified using coaxial RF amplifiers. The modulation strength $g$ can be tuned by changing the output voltage amplitude of the FPGA. We use a bias tee to apply an additional DC voltage to the modulator in the spin-down branch, to balance the optical lengths of the two polarization-maintaining branches when the modulation signals are absent. Inside the resonator we also incorporate an erbium-doped fiber amplifier (EDFA) to partially compensate for the round trip losses in the resonator, and a dense wavelength division multiplexing (DWDM) bandpass filter (from Fiberdyne Labs) to reduce the spontaneous emission noise from the EDFA. Finally, the light signal in the resonator is sampled through a $90:10$ fiber coupler, pre-amplified to improve the signal-to-noise ratio, and measured with a photodetector ($5~\text{GHz}$ bandwidth). The measured light intensity signal $\xi$ is a function of the laser frequency detuning $\delta\omega$, the roundtrip time $t$, and the input polarization state $\psi_\text{in}$.

A band structure measurement requires a recording of $\xi(\delta\omega, t)$ for a fixed input polarization $\psi_\text{in}$. In our setup, the laser frequency is swept slowly compared to the roundtrip time. We can therefore assume that the laser frequency detuning $\delta\omega$ remains the same within a single roundtrip time $T_R=250~\text{ns}$, but is slightly different for neighboring roundtrips~\cite{dutt2019experimental}. We estimate that the laser frequency difference between the neighboring roundtrips is $5\times10^{-4}\varOmega_R$. The light intensity recorded by the oscilloscope is truncated into pieces of length $T_R$, and different pieces are assigned different values of $\delta\omega$ in the $\xi(\delta\omega, t)$ plot. 

When performing tomographic reconstructions of the eigenstate, one needs to measure multiple copies of $\xi(\delta\omega, t, \psi_\text{in})$ with different input polarization states $\psi_\text{in}$. To achieve this, the polarization state generator is programmed to sequentially generate the polarization states $L$ (left-handed circular polarization), $H$ (horizontal polarization), $X$ ($+45^{\circ}$ linear polarization), $V$ (vertical polarization), $Y$ ($+135^{\circ}$ linear polarization), and $R$ (right-handed circular polarization). The switching period of the polarization state generator equals the time that the laser frequency is swept across one free spectral range $\varOmega_R$, which is approximately $500$ \textmu$\text{s}$. Thus, the measurement results within different free spectral ranges correspond to $\xi(\delta\omega, t, \psi_\text{in})$ with different input polarization states $\psi_\text{in}$.

We also perform the following calibrations prior to any acquisition of experimental data: 

\textbf{Calibrations of polarization controllers $U_1$ and $U_2$.} The birefringence of the single-mode fibers imposes additional rotations on the polarization state in the resonator, and needs to be considered when setting $U_1=\text{e}^{-i\frac{\pi}{4}\sigma_y}$ and $U_2=\sigma_z$. To calibrate for $U_1$, we inject the laser from point $P_1$ to the left instead of from collimator C1. Then the collimator C2 is replaced with a free-space polarimeter to determine the polarization rotation $U_1$. To calibrate for $U_2$, we inject known polarization states from collimator C1, and use a photodetector to measure the light intensity at point $P_2$ from the right. The polarization rotation $U_2$ can be determined based on the known input polarization state.

\textbf{Calibration of transmittance of the free-space beam splitter}, which can be polarization dependent and cause systematic errors in our tomographic eigenstate reconstructions. For different input polarization states, we use a photodetector to record the light intensity $\xi_{0}(\psi_\text{in})$ right after collimator C2. The experimentally measured signal $\xi(\delta\omega, t, \psi_\text{in})$ is later normalized with respect to $\xi_{0}(\psi_\text{in})$ for a particular input polarization.

\section*{Discussions on the lattice Hamiltonian}
\label{sec:modulation}

Our experimental setup can be modelled by the lattice Hamiltonian Eq.~(1) under approapriate assumptions. In Fig.~\ref{fig:S2}, the transmittances of the electro-optic modulators in the spin-up and spin-down branches are
\begin{align}
    \nonumber
    \tau_{\uparrow}(t) = 
    \begin{cases} \text{e}^{+igT_R \sin(\varOmega_R t)}, & 0 \leq t < T_R \\ \text{e}^{+igT_R \sin(M\varOmega_R t + \varphi)}, & T_R \leq t < 2T_R \end{cases} 
    \\
    \tau_{\downarrow}(t) = \begin{cases} \text{e}^{-igT_R \sin(\varOmega_R t)}, & 0 \leq t < T_R \\ \text{e}^{-igT_R \sin(M\varOmega_R t + \varphi)}, & T_R \leq t < 2T_R \end{cases}
    \label{eq:interleaving}
\end{align}
with $\tau_{\uparrow,\downarrow}(t+2T_R)=\tau_{\uparrow,\downarrow}(t)$. Here $g$ is the strength of the modulation, $T_R=2\pi/\varOmega_R$ is the roundtrip time for light propagation in the resonator, and $\varphi$ is the phase difference between modulation tones $\varOmega_R$ and $M\varOmega_R$. In the modulation signal we concatenate two sinusoidal waveforms of different frequencies. We refer to such modulation signals as an interleaving modulation scheme: frequencies in neighboring roundtrips are interleaved. We set the polarization rotations $U_1=\text{e}^{-i\frac{\pi}{4}\sigma_y}$ and $U_2=\sigma_z$.

We consider the polarization state at a location in the resonator, for example, after the collimator C2, and represent it by a Jones vector $\psi(T, t)$. Here $T$ is the slow-time variable representing the index of the roundtrip, and $t$ is the fast-time variable within each roundtrip. After two roundtrips, the polarization state evolves according to
\begin{equation}
    \psi(T+2T_R, t) = U_1
    \begin{bmatrix}
        \tau_{\uparrow}(t+T_R) & 0 \\
        0 & \tau_{\downarrow}(t+T_R)
    \end{bmatrix} U_2 U_1
    \begin{bmatrix}
        \tau_{\uparrow}(t) & 0 \\
        0 & \tau_{\downarrow}(t)
    \end{bmatrix} U_2 \psi(T, t).
    \label{eq:S7}
\end{equation}
In the weak modulation limit $g\ll\varOmega_R$, for $t\in[0, T_R)$, Eq.~(\ref{eq:S7}) can be approximated by
\begin{equation}
    \psi(T+2T_R, t) \approx \bigl[\sigma_0 + igT_R
                \bigl( \sin(\varOmega_R t) \sigma_z+ 
                       \sin(M \varOmega_R t + \varphi)\sigma_x\bigr)
                \bigr] \psi(T, t).
    \label{eq:S8}
\end{equation}
Approximating $\frac{\partial \psi(T, t)}{\partial T} \approx \bigl(\psi(T+2T_R, t) - \psi(T, t)\bigr) / \bigl(2T_R\bigr)$, one gets the following differential equation for $\psi$:
\begin{equation}
    i\frac{\partial \psi(T, t)}{\partial T} = -\frac{g}{2}
        [\sin(\varOmega_R t)\sigma_z + 
         \sin(M\varOmega_R t + \varphi)\sigma_x] \psi(T, t).
    \label{eq:S9}
\end{equation}
We then expand the polarization state on the basis of individual frequency modes in the resonator as $\psi(T, t) = \sum_{m} \psi_m(T) \text{e}^{i(\omega_0 + m\varOmega_R)t}$. Here $\omega_0$ is the frequency of the reference mode in the resonator, and $\psi_m(T)$ is the slowly-varying amplitude of the $m$-th frequency mode in the resonator. From Eq.~(\ref{eq:S9}), we derive
\begin{equation}
    i\frac{\text{d} \psi_m}{\text{d} T} = \frac{g}{4}
        (\text{e}^{i\frac{\pi}{2}\sigma_z}\psi_{m-1} + 
         \text{e}^{-i\frac{\pi}{2}\sigma_z}\psi_{m+1} +
         \text{e}^{i\frac{\pi}{2}\sigma_x}\text{e}^{i\varphi}\psi_{m-M} + \text{e}^{-i\frac{\pi}{2}\sigma_x}\text{e}^{-i\varphi}\psi_{m+M}).
    \label{eq:S10}
\end{equation}

We then map the frequency components to discretized lattice positions according to the following rule: given $m\in\mathbb{Z}$, we write $m = My + x$ with $x \in \{0, 1, ..., M-1\}$ and $y \in \mathbb{Z}$, and define $\psi_m = \psi_{x,y} \text{e}^{iy\varphi}$. This mapping is illustrated by the equivalence of Figs. 2(a) and 2(b). Eq.~(\ref{eq:S10}) then becomes a Schr\"{o}dinger-like equation that governs the evolution dynamics of a particle in a two-dimensional lattice:
\begin{equation}
    i\frac{\text{d} \psi_{x,y}}{\text{d} T} = \frac{g}{4}
    (\text{e}^{i\frac{\pi}{2}\sigma_z}\psi_{x-1,y} + 
     \text{e}^{-i\frac{\pi}{2}\sigma_z}\psi_{x+1,y} + 
     \text{e}^{i\frac{\pi}{2}\sigma_x}\psi_{x,y-1} + 
     \text{e}^{-i\frac{\pi}{2}\sigma_x}\psi_{x,y+1}).
    \label{eq:S11}
\end{equation}
Eq.~(\ref{eq:S11}) is subject to the twisted boundary condition that connects lattice sites $(0, y)$ and $(M-1, y-1)$: $\psi_{-1, y} = \psi_{M-1, y-1} \text{e}^{-i\varphi}$, $\psi_{M, y-1} = \psi_{0, y} \text{e}^{+i\varphi}$. The Hamiltonian associated with Eq.~(\ref{eq:S11}) is described by Eq.~(1), up to a constant factor, with non-Abelian gauge potentials $A_x = (\pi/2) \sigma_z$ and $A_y = (\pi/2) \sigma_x$. In other words, Eq.~(\ref{eq:S11}) is equivalent to $i(\text{d}{\psi}_{x,y} / \text{d}T) = \hat{H}\psi_{x,y}$, where $\hat{H}$ is the lattice Hamiltonian from Eq.~(1).

\section*{Methods of band structure measurements}
\label{sec:bandstructure}

In this section, we provide further details on the band structure measurements in our photonic synthetic frequency lattice. The technique is an extension of the two-dimensional band structure measurements reported in our previous work~\cite{cheng2023multi}: the resonance locations in the time-dependent transmission spectra correspond to the band energies along the sampling line $L_\varphi: k_y = Mk_x + \varphi~(\text{mod }2\pi)$ in the two-dimensional Brillouin zone. The linear band crossings at $k_x\in\{0, \pi\}$ revealed by the band structure measurements are one of the signatures of the non-Abelian lattice gauge fields.

We first describe the methods for simulating the band structure measurements. Such methods are used to generate Fig. 3(c), and are also essential in processing the experimental data in Fig. 3(d) as discussed below. Importantly, unlike in the previous section, here we do not make the assumption of weak modulation strength. For a given input polarization state $\psi_\text{in}$, the steady state in the resonator is a geometric series over all roundtrips:
\begin{align}
    \nonumber
    \psi_\text{ss}(\delta\omega, t) & = \sum_{m=0}^{\infty} \bigl[ U_1 \tau_Y(t) U_2 U_1 \tau_X(t) U_2 \text{e}^{2(i\delta\omega-\gamma_0/2)T_R} \bigr]^m
    \bigl[ U_1 \tau_Y(t) U_2 \text{e}^{(i\delta\omega-\gamma_0/2)T_R} \psi_\text{in} + \psi_\text{in} \bigr] \\
    & = \bigl[\sigma_0 - U_1 \tau_Y(t) U_2 U_1 \tau_X(t) U_2 \text{e}^{2(i\delta\omega-\gamma_0/2)T_R} \bigr]^{-1}
    \bigl[ \sigma_0 + U_1 \tau_Y(t) U_2 \text{e}^{(i\delta\omega-\gamma_0/2)T_R} \bigr] \psi_\text{in},
    \label{eq:S12}
\end{align}
where $\delta\omega$ is the frequency detuning of the laser, $\gamma_0$ represents the intrinsic loss rate in the resonator, and $\tau_X(t), \tau_Y(t)$ are polarization-dependent modulation signals in neighboring roundtrips: 
\begin{align}
    \tau_X (t) & = \begin{bmatrix}
        \text{e}^{+igT_R \sin(\varOmega_R t)} & 0 \\
        0 & \text{e}^{-igT_R \sin(\varOmega_R t)}
\end{bmatrix},~\text{and}\\
\tau_Y (t) &= \begin{bmatrix}
        \text{e}^{+igT_R \sin(M\varOmega_R t + \varphi)} & 0 \\
        0 & \text{e}^{-igT_R \sin(M\varOmega_R t + \varphi)}
\end{bmatrix}.
\end{align}
Eq.~(\ref{eq:S12}) is valid when the modulation frequency is $\varOmega_R$. Same as the previous section, here, $t\in[0, T_R)$ is the fast-time variable. For the sake of clarity, in the following we rewrite Eq.~(\ref{eq:S12}) as $\psi_\text{ss}(\delta\omega, t) = K^{-1}(\delta\omega, t) [\sigma_0 + Y(\delta\omega, t)] \psi_\text{in}$, with $K(\delta\omega, t) = \sigma_0 - U_1 \tau_Y(t) U_2 U_1 \tau_X(t) U_2 \text{e}^{2(i\delta\omega-\gamma_0/2)T_R}$ and $Y(\delta\omega, t) = U_1 \tau_Y(t) U_2 \text{e}^{(i\delta\omega-\gamma_0/2)T_R}$. Here $K^{-1}(\delta\omega, t)$ is the Green's function of the system, and the poles of $K^{-1}(\delta\omega, t)$ are associated with the resonances that match the band energies of the lattice Hamiltonian. The light signal directly observed by the photodetector, $\Tilde{\xi}(\delta\omega, t)$, is proportional to the steady-state intensity in the resonator:
\begin{align}
    \nonumber
    \Tilde{\xi}(\delta\omega, t) &= 
    \|\psi_\text{ss}(\delta\omega, t)\|^2 \\
    \nonumber
    &= \|K^{-1}(\delta\omega, t) [\sigma_0 + Y(\delta\omega, t)] \psi_\text{in}\|^2 \\
    \nonumber
    &= \|K^{-1}(\delta\omega, t) \psi_\text{in}\|^2 + 
       \|K^{-1}(\delta\omega, t) Y(\delta\omega, t) \psi_\text{in}\|^2 \\
    &+ \bigl(\psi_\text{in}^\dagger (K^\dagger(\delta\omega, t))^{-1} K^{-1}(\delta\omega, t) 
       Y(\delta\omega, t) \psi_\text{in} + \text{c.~c.}\bigr),
    \label{eq:S13}
\end{align}
where c.~c. stands for complex conjugate. In the last equals sign of Eq.~(\ref{eq:S13}), the first two terms correspond to resonances associated with polarization input $\psi_\text{in}$, and attenuated, rotated input $Y(\delta\omega, t) \psi_\text{in}$, respectively. Their resonance locations match the band energies. The interference terms in the last line of Eq.~(\ref{eq:S13}) are artifacts generated by the interleaving modulation scheme. To eliminate the last two terms, we calculate the following quantity:
\begin{equation}
    \dbtilde{\xi}(\delta\omega, t) = 
    \frac{1}{2} \bigl[
    \Tilde{\xi}(\delta\omega, t) + \Tilde{\xi}(\delta\omega + \varOmega_R/2, t) \bigr]
    = \|K^{-1}(\delta\omega, t) \psi_\text{in}\|^2 + 
      \|K^{-1}(\delta\omega, t) Y(\delta\omega, t) \psi_\text{in}\|^2.
    \label{eq:double-xi-1}
\end{equation}

The use of Eq.~(\ref{eq:double-xi-1}) allows us to determine the location of the resonances of the system without the artifacts of the interference terms in Eq.~(\ref{eq:S13}). In Fig. 3(c), we plot the simulation results of $\dbtilde{\xi}(\delta\omega, t)$ based on Eq.~(\ref{eq:double-xi-1}).

For experimental results, $\Tilde{\xi}(\delta\omega, t)$ is acquired directly from the readout of the photodetector in Fig.~\ref{fig:S2}, instead of evaluating $\|\psi_\text{ss}(\delta\omega, t)\|^2$ from Eq.~(\ref{eq:S12}). In Fig. 3(d), we calculate and plot $\dbtilde{\xi}(\delta\omega, t) = \bigl[\Tilde{\xi}(\delta\omega, t) + \Tilde{\xi}(\delta\omega + \varOmega_R/2, t) \bigr] / 2$ according to Eq.~(\ref{eq:double-xi-1}).

In Fig.~\ref{fig:S3}, we provide a comparison between $\Tilde{\xi}$ of Eq.~(\ref{eq:S13}) and $\dbtilde{\xi}$ of Eq.~(\ref{eq:double-xi-1}), both obtained experimentally. In $\Tilde{\xi}$, there are energy levels near both $\delta\omega = 0$ and $\delta\omega = 0.5 \Omega_R$. The existence of energy levels near $\delta\omega = 0.5 \Omega_R$ is due to the use of the interleaving modulation scheme as described in Eq.~(\ref{eq:interleaving}), where the modulation signal has a fundamental temporal period of $2T_R$. As defined in Eq.~(\ref{eq:double-xi-1}), $\dbtilde{\xi}$ has contributions from both the energy levels near $\delta\omega=0$ and $\delta\omega = 0.5 \Omega_R$. And $\dbtilde{\xi}$ is plotted in Fig. ~\ref{fig:S3}(c) as well as Fig. 3(d).

Finally, we notice that the band structure measurement results depend on the choice of input polarization state $\psi_\text{in}$. As shown in Figs. 3(c, d), the signal intensities at resonances are not uniform throughout the Brillouin zone. This arises due to the wavevector dependency   in the overlap between the eigenstate of the lattice Hamiltonian and the input polarization state -- which will be further discussed in the next section. Fig.~\ref{fig:S4} shows the simulation results of band structure measurements with various different input polarizations. The signal intensities at resonances are complementary for orthogonal input polarizations.

\section*{Methods of tomographic eigenstate reconstructions}
\label{sec:tomographic}
In this section, we provide more details on the tomographic reconstruction of eigenstates in photonic synthetic frequency dimensions. Using such methods, we are able to observe the reversal of the directions of eigenstate trajectories as the sampling lines move across Dirac points in the Brillouin zone. This is another signature of the non-Abelian lattice gauge fields.

\textbf{Framework of reconstruction of Stokes parameters of the eigenstates.} 
The steady-state polarization $\psi_\text{ss}(\delta\omega, \textbf{k}, \psi_\text{in})$ in the resonator is related to the input polarization $\psi_\text{in}$ by the Green’s function:
\begin{equation}
    \psi_\text{ss}(\delta\omega, \textbf{k}, \psi_\text{in}) = \frac{1}{\delta\omega - \hat{H}(\textbf{k}) + i\gamma_0/2} \psi_\text{in}.
    \label{eq:green}
\end{equation}
The Green's function here in Eq.~(\ref{eq:green}) corresponds to the definition of $K^{-1}(\delta\omega, t)$ in the weak-modulation limit, with $t$ corresponding to the wavevector $\textbf{k}$. We write $\hat{H}(\textbf{k}) = \omega_{-}(\textbf{k})\psi_{-}(\textbf{k})\psi_{-}^{\dagger}(\textbf{k}) + \omega_{+}(\textbf{k})\psi_{+}(\textbf{k})\psi_{+}^{\dagger}(\textbf{k})$ where $\omega_{\pm}(\textbf{k})$ is the energy of the upper/lower band and $\psi_{\pm}(\textbf{k})$ is the associated eigenstate, such that the steady state is a linear superposition of eigenstates:
\begin{equation}
    \psi_\text{ss}(\delta\omega, \textbf{k}, \psi_\text{in}) = \frac{\psi_{-}^{\dagger}(\textbf{k}) \psi_\text{in}}{\delta\omega-\omega_{-}(\textbf{k})+i\gamma_0/2} \psi_{-}(\textbf{k}) + \frac{\psi_{+}^{\dagger}(\textbf{k}) \psi_\text{in}}{\delta\omega-\omega_{+}(\textbf{k})+i\gamma_0/2} \psi_{+}(\textbf{k}).
    \label{eq:green-expand}
\end{equation}
For resonant conditions, for example, $\delta\omega = \omega_{-}(\textbf{k})$, we have
\begin{equation}
    \xi[\delta\omega=\omega_{-}(\textbf{k}),\textbf{k},\psi_\text{in}] = 
    \| \psi_\text{ss}(\delta\omega=\omega_{-}(\textbf{k}),\textbf{k},\psi_\text{in}) \|^2
    \approx \frac{4}{\gamma_0^2}|\psi_{-}^{\dagger}(\textbf{k})\psi_\text{in}|^2.
    \label{eq:resonant}
\end{equation}
The last approximation holds when $\gamma_0 \ll |\omega_{+}(\textbf{k}) - \omega_{-}(\textbf{k})|$. The steady state becomes dominated by eigenstate $\psi_{-}(\textbf{k})$. The intensity signal $\xi$ that one measures is the projection of the eigenstate onto the input state. The input state, produced by a polarization state generator, is reconfigurable and can be set as $\psi_\text{in} = H$ (horizontal), $V$ (vertical), $X$ ($+45^{\circ}$ linear polarized), $Y$ ($+135^{\circ}$ linear polarized), $L$ (left circular polarized), or $R$ (right circular polarized). The Stokes parameters of the eigenstate $\psi_{-}(\textbf{k})$ are thus calculated as the differences between two projective measurements:
\begin{align}
    \nonumber \langle\sigma_x\rangle(\textbf{k}) &= |\psi_{-}^{\dagger}(\textbf{k}) X|^2-|\psi_{-}^{\dagger}(\textbf{k}) Y|^2 \\
    \nonumber        &= \frac{\gamma_0^2}{4}[\xi(\delta\omega=\omega_{-}(\textbf{k}),\textbf{k},\psi_\text{in}=X) - \xi(\delta\omega=\omega_{-}(\textbf{k}),\textbf{k},\psi_\text{in}=Y)] \\ 
    \nonumber \langle\sigma_y\rangle(\textbf{k}) &= |\psi_{-}^{\dagger}(\textbf{k}) L|^2-|\psi_{-}^{\dagger}(\textbf{k}) R|^2 \\
    \nonumber        &= \frac{\gamma_0^2}{4}[\xi(\delta\omega=\omega_{-}(\textbf{k}),\textbf{k},\psi_\text{in}=L) - \xi(\delta\omega=\omega_{-}(\textbf{k}),\textbf{k},\psi_\text{in}=R)] \\
    \nonumber \langle\sigma_z\rangle(\textbf{k}) &= |\psi_{-}^{\dagger}(\textbf{k}) H|^2-|\psi_{-}^{\dagger}(\textbf{k}) V|^2 \\
                     &= \frac{\gamma_0^2}{4}[\xi(\delta\omega=\omega_{-}(\textbf{k}),\textbf{k},\psi_\text{in}=H) - \xi(\delta\omega=\omega_{-}(\textbf{k}),\textbf{k},\psi_\text{in}=V)].
                     \label{eq:stokes}
\end{align}

\textbf{Data processing in eigenstate tomographic reconstructions.} In the tomographic reconstruction algorithm described in the previous paragraph, data processing is required to accurately extract the eigenstate, for both simulations and experimental results in Fig. 4(b). In Eq.~(\ref{eq:double-xi-1}), we have introduced the quantity 
\begin{equation}
\dbtilde{\xi}(\delta\omega, t) = \|K^{-1}(\delta\omega, t) \psi_\text{in}\|^2 + \|K^{-1}(\delta\omega, t) Y(\delta\omega, t) \psi_\text{in}\|^2.
\end{equation}
The information encoded in $\dbtilde{\xi}$ involves contributions from input polarization states $\psi_\text{in}$ and $Y(\delta\omega, t) \psi_\text{in}$. The first term $\xi(\delta\omega, t, \psi_\text{in}) = \|K^{-1}(\delta\omega, t) \psi_\text{in}\|^2$ needs to be extracted and put into Eq.~(\ref{eq:stokes}) for proper eigenstate reconstruction. To achieve this, we consider the steady-state signal at the time point $t+T_R \in [T_R, 2T_R)$, when the modulation frequency is $M\varOmega_R$. Following similar calculations as in Eqs.~(\ref{eq:S12})–(\ref{eq:double-xi-1}), we get
\begin{equation}
    \dbtilde{\xi}(\delta\omega, t+T_R)
    = \text{e}^{-\gamma_0T_R}
      \|K^{-1}(\delta\omega, t) \psi_\text{in}\|^2
    + \text{e}^{+\gamma_0T_R}
      \|K^{-1}(\delta\omega, t) Y(\delta\omega, t) \psi_\text{in}\|^2.
    \label{eq:double-xi-2}
\end{equation}
Combining Eqs.~(\ref{eq:double-xi-1}) and~(\ref{eq:double-xi-2}), we can solve for the desired quantity $\xi(\delta\omega, t, \psi_\text{in}) = \|K^{-1}(\delta\omega, t) \psi_\text{in}\|^2$. In an experiment, the intrinsic loss rate of the resonator $\gamma_0$ can be estimated from the linewidth of the resonances in the measured spectra.

\end{document}